\documentstyle [12pt]{article}
\newcommand{\be}{\begin{equation}}
\newcommand{\ee}{\end{equation}}
\newcommand{\bea}{\begin{eqnarray}}
\newcommand{\ena}{\end{eqnarray}}

\def\Im   {\mbox{Im}}
\def\Re   {\mbox{Re}}

\def\(    {\left( }    \def\)   {\right) }
\def\[    {\left[}    \def\]   {\right] }
\def\<    {\big{\langle}}  \def\> {\big{\rangle}}
\def\txt #1 {\qquad {\mbox{#1}} \qquad}
\def\lsim{\; \raise0.3ex\hbox{$<$\kern-0.75em\raise-1.1ex\hbox{$\sim$}}\; }
\def\gsim{\; \raise0.3ex\hbox{$>$\kern-0.75em\raise-1.1ex\hbox{$\sim$}}\; }
\def\@versim#1#2{\lower0.2ex\vbox{\baselineskip\z@skip\lineskip\z@skip
  \lineskiplimit\z@\ialign{$\m@th#1\hfil##\hfil$\crcr#2\crcr\sim\crcr}}}

\begin{document}
\begin{titlepage}
\begin{center}
\rightline{CERN-TH.7271/94}
\rightline{KSUCNR-008-94}
\bigskip
\bigskip
\bigskip
\bigskip

{\bf {PROBLEMS OF PERTURBATION SERIES}}\\
{\bf {IN NON-EQUILIBRIUM QUANTUM FIELD THEORIES}}\\
\bigskip
\vspace{0.8cm}
{\large T. Altherr$^{1,2}$ and D.~Seibert$^3$}\\
\bigskip
{\em 1) Theory Division, CERN, CH-1211 Geneva 23, Switzerland}\\
{\em 2) L.A.P.P., BP110, F-74941 Annecy-le-Vieux Cedex, France}\\
{\em 3) Physics Department, Kent State University, Kent, OH 44242, USA}
\end{center}
\medskip

\centerline{ \bf{Abstract}}
In the standard framework of non-equilibrium quantum field theories,
the pinch singularities associated to multiple products of
$\delta$-functions do not cancel in a perturbative expansion unless the
particle distributions are  those for a system in thermal and
chemical equilibrium.
 \bigskip

\vfill

\leftline{CERN-TH.7271/94}
\leftline{May 94}
\end{titlepage}

\setcounter{footnote}{0}
Since the pioneering work of Keldysh \cite{Kel}, there have been many
attempts to develop a formalism for non-equilibrium systems. Most of the
studies follow the original method of Schwinger \cite{Sch}, which uses a
Closed-Time-Path (CTP) in the complex-time plane \cite{CTP}. Unlike the
imaginary-time formalism, but closely related to the real-time formalism,
this approach leads to a $2\times 2$ matrix stucture for the propagator
\cite{LvW}.

One other approach, which is more recent, is  Thermo-Field Dynamics
\cite{TFD}. There, the doubling of the degrees of freedom is assumed from the
beginning. For equilibrium systems, one can show that the theory is completely
equivalent to the real-time formalism \cite{LvW}. The non-equilibrium version
of TFD has been intensively studied over the last few years \cite{TFDNE}.

In this letter, we shall not discuss the relative advantages of one method
over the other, but shall just make a very simple remark about the
perturbative expansion, when one is using non-equilibrium particle
distributions.  We consider the scalar case. The most common form of the
free propagator that is used in the literature is \cite{CTP,TFDNE}
\begin{eqnarray}
 \( \begin{array}{cc} D_{11}(K) & D_{12}(K) \\
                      D_{21}(K) & D_{22}(K) \end{array} \)
       &=&   \(
           \begin{array}{cc}
 \Delta(K) & 0 \\
 0         & \Delta^*(K)
           \end{array} \) \nonumber\\
          &+& \(
           \begin{array}{cc}
  n(k)               & \theta(k_0)+n(k)  \\
  \theta(-k_0)+n(k)  & n(k)
           \end{array} \) 2\pi\delta(K^2-m^2)
,\label{PROP}\end{eqnarray}
with the usual vacuum Feynman propagator
\be
\Delta(K) = {i\over K^2 - m^2 + i\epsilon}
,\ee
and where $n(k)$ is an arbitrary distribution function (positive-definite).
As a matter of fact, this is the only difference with the equilibrium
case, where the propagator is exactly the same as in eq.~(\ref{PROP}), except
that $n(k)=1/(e^{|k_0|/T}-1)$.

Self-interactions are taken as in the equilibrium case, i.e. one
needs to distinguish two types of vertices, which will be called type-1 and
type-2 vertices in the following. For a Feynman diagram with a certain
configuration of type-1 and type-2 vertices, one uses a $D_{ab}(K)$
propagator when the momentum $K$ flows from a type-$a$ vertex to a
type-$b$ one. For a $g^n\phi^n/n!$ interaction, type-1 vertices get a factor
$(ig)$, whereas type-2 vertices get an opposite factor, $(-ig)$.
These Feynman rules set up a formalism for non-equilibrium systems. They have
been used many times in the literature [1--9].
 \bigskip

Let us now see if the above Feynman rules give
a well-defined perturbative expansion. Knowing the problems which arise at
equilibrium, the first obvious question is the absence of pathologies, or
``pinch singularities''. This is due to the presence of several
$\delta(K^2-m^2)$ terms  in eq.~(1). They appear as multiple
products in high order calculations, and for individual graphs, lead to
mathematically ill-defined expressions. At equilibrium, one has to sum over the
different types of vertices in order to obtain the cancellation of such
pathologies \cite{LvW}. Surprisingly, this very question has  almost
always been occulted for the non-equilibrium case \cite{BP,Wel1}.

We first consider the simplest case where such a cancellation should occur, if
any. Suppose that we want to calculate the tadpole contribution at the 2-loop
level shown in fig.~1, in the $g(\phi^4)_4$ interaction model. It is given by
\be
g^2\int {d^4P\over (2\pi)^4} \int {d^4K\over (2\pi)^4}
\[ (D_{11}(P))^2 D_{11}(K) - D_{12}(P) D_{21}(P) D_{22}(K) \]
.\ee
Observing that the tadpole does not have any imaginary part and  that $\Re
D_{11}(K)=\Re D_{22}(K)$, one can factorize the $K$-integral and obtain
\be
g^2 \int {d^4K\over (2\pi)^4} \Re D_{11}(K) \int {d^4P\over (2\pi)^4}
\[ (D_{11}(P))^2  - D_{12}(P) D_{21}(P)  \]
.\ee
Then an easy algebraic manipulation shows that the term in the square
brackets simplifies to
\be
\[ ...\] = (\Delta(P))^2 + n(p) \( (\Delta(P))^2 - (\Delta^*(P))^2\)
,\ee
which shows the absence of pinch singularities, or $\Delta(P)\Delta^*(P)$ terms
in the final result. The cancellation procedure works essentially the same
way as in the equilibrium case and is here totally independent of the
distribution function $n(p)$. This result had previously been found in
\cite{BP}. \bigskip

One should not claim victory too soon, though. This simple exercise in
the $\phi^4$ model {\it does not} illustrate well the game of cancellation
that is at play. In particular, in the previous example only two terms are
involved, although the general case involves four terms.

Let us next consider a more complicated case. The only
place where the pinch singularities can occur is in repeated self-energy
insertions. Consider for instance the diagram shown in fig.~2.
It contains the following expression
\bea
\sum_{a,b}  D_{1a}(P)\Sigma_{ab}(P)D_{b2}(P) =
  D_{11}(P)\Sigma_{11}(P)D_{12}(P) +
D_{11}(P)\Sigma_{12}(P)D_{22}(P)\nonumber\\
+ D_{12}(P)\Sigma_{21}(P)D_{12}(P) + D_{12}(P)\Sigma_{22}(P)D_{22}(P)
,\label{DYSON}\ena
which must be free of pinch singularities, as it enters directly, at the
two-loop level, into the calculation of a physical quantity (the decay rate),
as
we shall see in the following.
For the same reason, the cancellation must also take place separately
for $p_0>0$ and for $p_0<0$.

 The different components of the
self-energy can be related to each other by \bea
        \Sigma_{11}(P) &=& - \Sigma_{22}^*(P)  \nonumber\\
        \Im\Sigma_{11}(P) &=& {i\over 2} (\Sigma_{12}(P) + \Sigma_{21}(P))
.\label{2POINT}\ena
These relations follow from the definition of the two-point matrix Green's
function in different chronological products, using the standard CTP contour
\cite{CTP}. They are independent of perturbation theory.  Then, using
(\ref{2POINT}), one can show that eq.~(\ref{DYSON}) is free of pinch
singularities provided
\be
\( \theta(p_0)n(p) - \theta(-p_0)(1+n(p))\) \Sigma_{12}(P)
=\epsilon(p_0) (\theta(p_0)+n(p))\Sigma_{21}(P)
.\label{KMSNE}\ee
For $p_0 > 0$, one has
\be
 n(p)\Sigma_{12}(P) = (1+n(p))\Sigma_{21}(P)
,\ee
and when $n(p)=1/(e^{p_0/T} -1)$, this gives
\be
\Sigma_{12}(P) = e^{p_0/T} \Sigma_{21}(P)
,\ee
which shows that eq. (\ref{KMSNE}) can in fact be regarded as a non-equilibrium
extension of the KMS relation.

The quantities $\Sigma_{12}$ and $\Sigma_{21}$, which are related through
eq.~(\ref{KMSNE}), are two independent physical quantities.
For a $\lambda\phi^3$ interaction model in $n$ space-time dimensions,
one has, at one loop,
 \begin{eqnarray}
-i\Sigma_{12}(P) &=& \lambda^2\int {d^n K\over (2\pi)^{n-2}}
(\theta(k_0)+n(k)) (\theta(p_0-k_0)+n(p-k)) \nonumber \\
&&\delta(K^2-m^2)\delta((P-K)^2-m^2)
.\label{S12}\end{eqnarray}
The kinematics are the same as in the equilibrium case, i.e. $(E-p)/2 \le
k_0 \le (E+p)/2$, for $p_0=E\ge 0$ and $P^2\ge m^2$. Then, the statistical
factors in the above equation are just the ones corresponding to outgoing
particles. As in the equilibrium case, this allows us to relate $\Sigma_{12}$
with the absorption (or decay) rate of the particle \cite{Wel2}. Similarly
 \begin{eqnarray} -i\Sigma_{21}(P) &=& \lambda^2\int {d^n K\over
(2\pi)^{n-2}}  (\theta(-k_0)+n(k))
(\theta(-p_0+k_0)+n(p-k)) \nonumber \\ &&\delta(K^2-m^2)\delta((P-K)^2-m^2)
\label{S21}\end{eqnarray}
is related to the  emission (or creation) rate.
Strictly speaking, when $P$ is on shell, there is no kinematical phase space
anymore in eqs.~(\ref{S12}) and (\ref{S21}) and both expressions are equal
to zero. But this is only true at the one-loop level. Beyond one-loop
calculations, $\Sigma_{12}$ and $\Sigma_{21}$ no longer vanish when the
external momentum goes on shell. Another way of looking at the problem is to
consider two types of particles, $\phi_1$ and $\phi_2$, with two different
masses
and a coupling such as $\lambda \phi_1\phi_2^2$ (for details see \cite{Wel2}).
In this case, the interpretation of $\Sigma_{12}$ and $\Sigma_{21}$ as,
respectively, the decay and emission rates, is more transparent.

According to the above discussion, the time evolution of the particle number
density follows \cite{Wel2}
 \be
-2ip_0 {d n(p,t)\over dt}=  (1+n(p))\Sigma_{21}(P) - n(p)\Sigma_{12}(P)
.\ee
We see that in order to have the cancellation of pinch singularities, one must
have
 \be
{d n(p,t)\over dt}=  0
,\ee
which is quite disappointing for a non-equilibrium framework~!
This is the first contradiction. However, this is not a real problem as it can
be realized that for the propagator defined in (1), it must be assumed that the
time variation of the density matrix is slow compared with the typical time
scale between the particle interactions. If this is not the case, the Fourier
transform in (1) just does not make sense. In the context of using (1),
there is reversibility at the microscopic level, but one can impose some
irreversibility at the macroscopic level.

But the main trouble is that, even assuming a slow variation of the density
matrix, eq.~(\ref{KMSNE}) is not guaranteed to hold. The point is that the
micro-reversibility conditions are well-known to be satisfied only by {\it
equilibrium} distributions \cite{KLN}. This can be verified by explicit
calculations using eqs. (\ref{S12}) and (\ref{S21}). {\it
Equation~(\ref{KMSNE}) can only be satisfied if the distribution $n(p)$ is of
the Bose--Einstein type}. The only alternative to this result is to give up
energy
conservation at the vertices, which is not  very satisfactory.

To see how deeply rooted the problem is, let us consider again the case of two
types of particles, with two different initial temperatures $T_1$ and $T_2$,
and
a single weak interaction $\lambda\phi_1\phi_2^2$, which is switched on at some
arbitrary time $t_0$. Then the free propagators for $\phi_1$ and
$\phi_2$ are just the same as in eq.~(1), with $n(p)$ the
Bose--Einstein distribution, but different temperatures for $\phi_1$
and $\phi_2$. At one loop, the self-energy for $\phi_1$ involves only
$\phi_2$ fields, and it obeys the relation
 \be \Sigma_{12}(P) =
e^{p_0/T_2} \Sigma_{21}(P) ,\ee as $\phi_2$ is thermalized with temperature
$T_2$. On the other hand, the temperature that  enters
$n(p)$ in eq.~(\ref{KMSNE}) is $T_1$, not $T_2$, so that the cancellation of
pathologies occurs only when $T_1=T_2$.

Perhaps the most disappointing lesson of all this is to realize that it is
not even possible to look at small deviations from equilibrium.
Also, this problem is not specific to the
relativistic case: it shows up in the same fashion in the non-relativistic
limit.
 \bigskip

Except for \cite{Wel1}, this fact seems to have gone unnoticed in the
literature. One way of solving this problem is to use Schwinger--Dyson
equations. By using only exact propagators in one-loop calculations, there is
no possibility of having pinch singularities.  This is equivalent to giving up
perturbation theory. As a matter of fact, this has been the usual way of doing
calculations for non-equilibrium systems \cite{CTP}. However, at some point,
and unless the problem can be solved exactly, which is rather rare, one is
forced to use some perturbative input. In the light of our results, this has
very
weak justification as the bare perturbation series is ill-defined. In
particular,
terms that are associated with pinch singularities (which can be regularized by
introducing some finite width in the propagator) are likely to give large and
uncontrollable contributions.
\bigskip

In conclusion, it is impossible to make use of perturbation series with the
propagators (1) outside an equilibrium framework. One must use
time-independent Bose--Einstein distribution functions (the same is true for
fermion fields, which have to obey the Fermi--Dirac statistics). This
guarantees the cancellation of pinch singularities at all orders of the
perturbation series.  This cancellation is intimately tied to the
micro-reversibility conditions. Note also that the same conditions ensure the
cancellation of infrared and mass singularities (KLN theorem), which defines
a well-behaved perturbation series \cite{KLN}.

The problem can clearly be solved in principle by working in the $T=0$
representation of the system, but then the calculation is computationally
intractable because of the complexity of the system.  If we choose to work
with any fixed-$T$ representation, then the system quickly leaves the vacuum
state.  We cannot treat the problem using the propagators (1), because then
the $\delta$-function pathologies discussed here arise, so our only choice
is to allow the state of the system to depart from the thermal vacuum;
again, the problem quickly becomes computationally intractable because of
the complexity of the state.  Finally, we could choose the closest finite-$T$
vacuum to approximate the state of the system at each time, but then we are
faced with the intractable task of transforming the state of the system as a
function of time as the Fock states change with the changing thermal vacua.

We stress again that anything perturbative beyond linear response theory
does not seem easily realizable. At present, there does not exist any correct
way for deriving a consistent perturbative Green's function formalism for a
system that is even slightly out of equilibrium, without losing all of the
advantages of working in finite-$T$ vacua.

\section*{Acknowledgements}

The work of D.S. was supported in part by the U.S. Department of Energy
under Grant\\ No.\ DOE/DE-FG02-86ER-40251.

\newpage
{\large\bf Figure captions\\[0.5cm]}
\begin{description}
\item[Fig.~1] A 2-loop contribution to the self-energy.
\item[Fig.~2] A particular summation of self-energy insertions.
\end{description}

 \end{document}